\documentclass[a4paper,twocolumn,11pt,accepted=2022-07-05]{quantumarticle}
\pdfoutput=1
\usepackage{bbm,amsmath,amssymb, url,graphicx,fullpage,amsthm,physics}
\usepackage[colorlinks,pdftex]{hyperref} 
\usepackage[dvipsnames]{xcolor}
\usepackage[utf8]{inputenc}
\usepackage[english]{babel}
\usepackage[T1]{fontenc}

\usepackage{cite}          

\usepackage{hyperref} 

\usepackage{tikz}
\usepackage{lipsum}

\begin{document}

\title{Classical Simulation of High Temperature Quantum Ising Models}

\author{Elizabeth Crosson}
\affiliation{Phasecraft Inc., Washington DC, USA}

\author{Sam Slezak}
\affiliation{Information Sciences, Los Alamos National Laboratory, Los Alamos, NM, USA }
\affiliation{Center for Quantum Information and Control, University of New Mexico, Albuquerque, NM 87131, USA}

\maketitle

\begin{abstract}
  We consider generalized quantum Ising models, including those which could describe disordered materials or quantum annealers, and we prove that for all temperatures above a system-size independent threshold the path integral Monte Carlo method based on worldline heat-bath updates always mixes to stationarity in time $\mathcal{O}(n \log n)$ for an $n$ qubit system, and therefore provides a fully polynomial-time approximation scheme  for the partition function.   This result holds whenever the temperature is greater than four plus twice the maximum interaction degree (valence) over all qubits, measured in units of the local coupling strength.  For example, this implies that the classical simulation of the thermal state of a superconducting device modeling a frustrated quantum Ising model with maximum valence of 6 and coupling strengths of 1 GHz is always possible at temperatures above 800 mK. Despite the quantum system being at high temperature, the classical spin system resulting from the quantum-to-classical mapping contains strong couplings which cause the single-site Glauber dynamics to mix slowly, therefore this result depends on the use of worldline updates (which are a form of cluster updates that can be implemented efficiently).  This result places definite constraints on the temperatures required for a quantum advantage in analog quantum simulation with various NISQ devices based on equilibrium states of quantum Ising models.  
\end{abstract}

\section{Introduction}
Quantum transverse Ising models (TIM) have occupied a distinguished role in the study of many-body quantum systems~\cite{sachdev2007quantum, suzuki2012quantum}.   The 1D TIM has been extensively studied as an exactly solvable model, which exemplifies statistical mechanical dualities by its relation to free spinless fermions and to the 2D classical Ising model~\cite{RevModPhys.36.856}.   In Hamiltonian complexity, the ground state problem for the TIM is complete for the class StoqMA that is on the border of quantum and classical complexity~\cite{bravyi2017complexity,cubitt2018universal} and are universal for a broad class of stoquastic adiabatic computations~\cite{albash2018adiabatic}.  Effective Ising interactions are also ubiquitous~\cite{schauss2018quantum,zhang2017observation} in NISQ era~\cite{preskill2018quantum} devices.  A general TIM on $n$-qubits has the form
\begin{equation}
H = \sum_{i\sim j} a_{ij} Z_i Z_j + \sum_i b_i Z_i - \sum_i \Gamma_i X_i \label{eq:genTIM},
\end{equation}
where the couplings $\{a_{ij}\}, \{b_i\},\{\Gamma_i\}$ are all real. Here $i \sim j$ denotes adjacency in the interaction graph which associates qubits with vertices and pairwise Hamiltonian terms with edges.  

Every Hamiltonian of the form \eqref{eq:genTIM} is \emph{stoquastic}~\cite{bravyi2006merlin}, which means that there is some choice of local basis in which all of the off-diagonal matrix elements of $H$ are real and non-positive.  The computational basis matrix elements of $H$ satisfy the required property after conjugating $H$ by the 1-local unitary $\otimes_{i=1}^n Z^{\frac{1}{2}(1-\textrm{sign}(\Gamma_i))}$ (this unitary is said to ``cure the sign problem''~\cite{marvian2019computational,klassen2019two}), and so we take $\Gamma_i > 0$ for each $i$ without loss of generality.  Approximating the ground energy of a stoquastic local Hamiltonian problem can be done in the complexity class AM~\cite{bravyi2006merlin} while for general local Hamiltonians it is QMA-complete~\cite{kempe2004complexity}.   The special case of frustration-free stoquastic adiabatic computation can be classically simulated in polynomial time~\cite{bravyi2009complexity} (though this result does not include any non-trivial transverse Ising models due to frustration caused by the anticommuting nature of Pauli $X$ and $Z$), and the random walk used in that result has recently been applied to show that quantum probabilistically checkable proofs based on reductions that preserve the stoquastic property would imply $\textrm{MA} = \textrm{NP}$~\cite{aharonov2019stoquastic}.   TIM of the form \eqref{eq:genTIM} with polynomially bounded coupling strengths are universal for bounded-degree stoquastic adiabatic computation~\cite{bravyi2017complexity, cubitt2018universal}.  

In 1977 Suzuki introduced a Markov chain Monte Carlo algorithm for approximating the partition function of quantum Ising models and other stoquastic Hamiltonians~\cite{suzuki1977monte} (this algorithm motivates special consideration models with restrictions on the signs of Hamiltonian matrix elements).  Suzuki's method, which is now called path integral Monte Carlo (PIMC), is based on relating the quantum partition function of interest to a partition function of a classical spin system~\cite{suzuki1976relationship}, and using a Markov chain Monte Carlo procedure~\cite{levin2017markov} to approximate properties of the latter.  In the last decade rigorous polynomial-time upper bounds on the run time of Suzuki's algorithm been obtained for 1D systems with power-law interactions at constant temperature~\cite{crosson2018rapid}, specific problems related to quantum annealing~\cite{crosson2016simulated,jiang2017scaling,jarret2016adiabatic}, and for ferromagnetic systems on arbitrary graphs for temperatures which are at least inverse polynomial small~\cite{bravyi2015monte, bravyi2017polynomial}.  These examples all adopt premises that preclude computational complexity obstructions to finding a fully polynomial-time approximation scheme (FPRAS) for the partition function.  In contrast, for general low-temperature classical Ising systems with non-ferromagnetic interactions there can be no FPRAS for the partition function unless randomized polynomial-time is equal to NP~\cite{jerrum1993polynomial}.   In the present work we treat arbitrary non-ferromagnetic interactions, but restrict the temperature to be sufficiently high that no complexity obstructions can occur.  Algorithms for simulating high temperature quantum systems have been a subject of recent interest~\cite{harrow2019classical,kuwahara2019clustering}, with those works finding complementary domains of simulation to the algorithm presented here.

\noindent \textbf{Main result.} We establish the existence of a temperature threshold such that the PIMC method is guaranteed to yield an FPRAS for the partition function $\mathcal{Z} \equiv \textrm{tr}\left(e^{-\beta H}\right)$ at any temperature above that threshold.   For the same range of temperatures one can also use PIMC to approximately sample from the distribution $\mu_\beta(z) \equiv \langle z | e^{-\beta H}|z\rangle/\mathcal{Z}$ of computational basis measurements ($z \in \{0,1\}^n$) with total variation distance error $\epsilon$ after $\mathcal{O}(n \log(n/\epsilon))$ heat-bath worldline updates.  The runtime bound of $\mathcal{O}(n \log n)$ is optimal for a Markov chain method with updates that act locally on the (classical degrees of freedom associated with the) qubits~\cite{hayes2005general}. 

In terms of the maximum coupling strength $J \equiv \max_{ij} |a_{ij}|$ and maximum interaction degree $\Delta \equiv \max_i |\{j :  |a_{ij}| \neq 0\}$ these results hold whenever the inverse temperature satisfies
\begin{equation}\label{eq:thresholdTemp}
\beta \leq \frac{1}{2 J(\Delta + 2)}.
\end{equation}
This result is derived from an analysis of the mixing time of the PIMC Markov chain with worldline updates with heat-bath transition probabilities.  We show that the mixing time of this Markov chain is $\mathcal{O}(n \log n)$ when \eqref{eq:thresholdTemp} is satisfied. Therefore the overall sampling algorithm runs in time $\mathcal{O}(R n \log n)$ where $R$ is the time it takes to perform a single heat-bath worldline update.   In appendix \ref{sec:updates} We analyze a standard implementation of these updates based on the cavity method~\cite{krzakala2008path} to show that $R = \mathcal{O}(\beta \log n)$.  This algorithm appeals to the continuous imaginary-time limit of the quantum-to-classical mapping to obtain a run time that is independent of the Trotter number, and depends only on the expected number of jumps along the imaginary-time direction.  In appendix \ref{sec:pfail} maximum number of these jumps is determined by a Poisson process with mean $\mathcal{O}(\beta)$.

\section{Preliminaries}
\paragraph{Path Integral Monte Carlo.} The PIMC method is based on Suzuki's quantum-to-classical mapping~\cite{suzuki1976relationship} from a system of $n$ qubits to a system of $L \times n$ classical spins, which are sometimes described as $L$ ``replicas'' of the original system that are coupled together ferromagnetically.   For classical configurations $\mathbf{z} \in \Omega$ we either write $\mathbf{z} \equiv (z_1,...,z_L)$ with $z_i \in \{-1,1\}^n$ or $\mathbf{z} \equiv [\bar{z}_1,...,\bar{z}_n]$ where $\bar{z}_i \in \{-1,1\}^L$.  Individual spins are denoted by $z_{ij} \in\{-1,1\}$ where $i \in \{1,...,L\}$ and $j \in \{1,...,n\}$, and we may write $z_{ij} \in \bar{z}_i$ or $z_{ij} \in z_i$.   In the form $\mathbf{z} = (z_1,...,z_L)$ the $z_i$ are called ``replicas'' or ``(imaginary) time slices'', while in the form $\mathbf{z} = [\bar{z}_1,...,\bar{z}_n]$ the $\bar{z}_i$ are called ``worldlines.''    The goal of the PIMC method is to sample from the following equilibrium distribution on the classical spins,
\begin{equation}\label{eq:measure}
\pi(\mathbf{z}) \equiv \frac{1}{Z} e^{ -\frac{\beta}{L} \sum\limits_{i = 1}^L \sum\limits_{j\sim k}a_{jk}z_{ij}z_{ik}+\sum\limits_{j} b_jz_{ij}} \prod_{i=1}^n \phi(\bar{z}_i)
\end{equation}
where $Z$ is proportional to $\mathcal{Z}$ and 
$$\phi(\bar{z}_i) \equiv \tanh\left(\frac{\beta\Gamma_j}{L}\right)^{|\{k:z_{jk}\neq z_{j(k+1)}|}.$$  The distribution $\mu_\beta$ is the marginal distribution of $\pi$ on a single replica.  See \cite{bravyi2015monte} for a full derivation of the PIMC method.

The distribution \ref{eq:measure} is sampled by generalized heat-bath updates~\cite{dyer2004mixing} (i.e. sampling a region of spins from the conditional distribution that fixes spins outside of that region) applied to one worldline at a time, which are called worldline heat-bath updates.  If two configurations $\mathbf{z} = [\bar{z}_1,...,\bar{z}_j,...,\bar{z}_n]$ and $\mathbf{z}' = [\bar{z}_1,...,\bar{z}_j',...,\bar{z}_n]$ differ only in the $j$-th worldline, then the transition probability $P(\mathbf{z}, \mathbf{z}')$ is
\begin{align*}
P(\mathbf{z}, \mathbf{z}') \equiv&\frac{1}{n}  \frac{\pi([\bar{z}_1,...,\bar{z}_j',...,\bar{z}_n])}{\sum_{z_j''}\pi([\bar{z}_1,...,\bar{z}_j'',...,\bar{z}_n])} \\ \equiv&\frac{1}{n} \pi_j(\bar{z}_j'|\mathbf{z})
\end{align*}
where the factor of $n^{-1}$ is the probability of selecting worldline $j$.  Our proof makes use of the form
\begin{equation}\label{eq:conddistrubution}
    \pi_j(\bar{z}_j'|\mathbf{z}) \equiv  \frac{e^{-\frac{\beta}{L}g_j(\bar{z}_j'|\mathbf{z})}\phi(\bar{z}_j')}{\sum_{\bar{z}_j''}e^{-\frac{\beta}{L}g_j(\bar{z}_j''|\mathbf{z})}\phi(\bar{z}_j'')}
\end{equation}
where the conditional energy function $g_j(\bar{z}_j'|\mathbf{z})$ is
$$ 
g_j(\bar{z}_j'|\mathbf{z}) \equiv \sum_{k=1}^L\left (\sum_{i\in\mathcal{N}(j)}a_{ij}z_{ki}z_{kj}' + \sum_j b_jz_{kj}'\right),
$$
with $\mathcal{N}(j) \equiv \{i: a_{ij} \neq 0\}$.

\begin{figure}[htbp]
    \centering
    \includegraphics[width=1.0\linewidth]{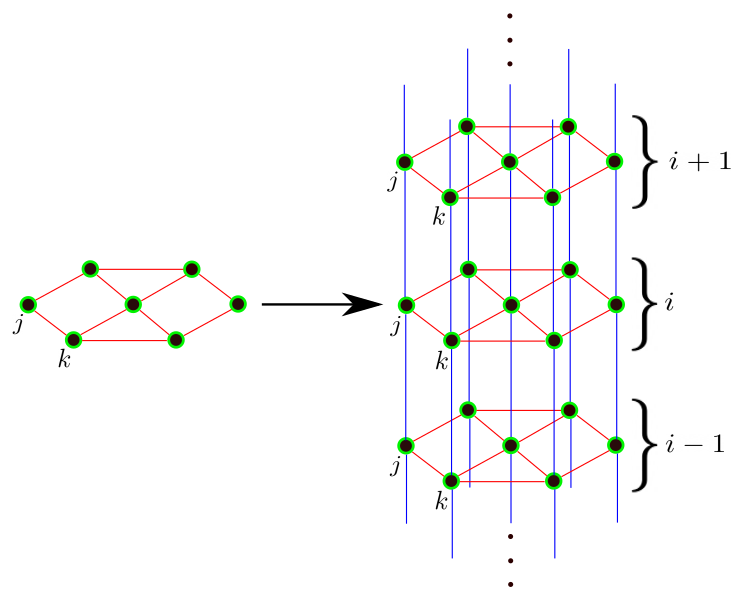}
    \caption{
        A visual representation of the path integral Monte Carlo quantum to classical mapping. A quantum system of seven qubits on the left with the $ZZ$ couplings represented by red lines is transformed into the classical system on the right with the $a_{jk}z_{ij}z_{ik}$ Ising interactions represented by red lines, and the $\phi(\bar z_i)$ worldline interactions represented by the blue lines.
    }
    \label{fig:Q-to-C-mapping}
\end{figure}

\paragraph{Mixing times and path coupling.} Given a Markov chain with stationary distribution $\pi$, transition matrix $P$, and state space $\Omega$, let $P^t(x,\cdot)$ be the distribution that results from starting at the initial state $x \in \Omega$ and evolving for $t$ steps.   We measure the distance from stationarity after $t$ steps as the total variation distance between $P^t(x,\cdot)$ and $\pi$ for the worst-case initial state,
\begin{equation}
d(t) \equiv \max_{x \in \Omega}\|P^t(x,\cdot) - \pi\|_{\textrm{TV}} 
\end{equation}
and the mixing time of the chain is
\begin{equation}
t_\textrm{mix}(\epsilon) \equiv \min_t\{t : d(t') < \epsilon \textrm{ for all } t' \geq t\}.
\end{equation}
The mixing time is an appropriate notion of convergence in this setting because the total variation distance also bounds the difference in expectation values of observables.

A powerful and versatile technique for bounding the mixing time of Markov chains is based on the notion of a coupling.  A coupling of two probability distributions $\mu$ and $\nu$ is a pair of random variables $(X,Y)$ defined on the same probability space with $X$ distributed according to $\mu$ and $Y$ distributed according to $\nu$.   In the analysis of mixing we seek to define a coupling $(X_t,Y_t)$ where $X_t$ is distributed according to $P^t(x,\cdot)$ and $Y$ is distributed according to $P^t(y,\cdot)$, thereby analyzing two trajectories of the Markov chain starting from distinct initial states.   If for each $x,y \in \Omega$ we have such a coupling $(X_t,Y_t)$ with $X_0 = x$ and $Y_0 = y$, then the time $\tau = \min \{t : X_t = Y_t\}$ it takes for the two copies of the chain to coincide can be used to upper bound the distance from stationarity,
\begin{equation}
d(t) \leq \max_{x,y \in \Omega} \mathbf{P}_{x,y} \left \{\tau \geq t \right \}
\end{equation}
In other words, the distance to stationarity at time $t$ is upper bounded by the probability that the two branches of the coupling have not coincided yet at time $t$, for the worst-case possible pair of starting states.  In applications of this method one uses the fact that the two branches of the coupling are defined on the same probability space (which, for the sake of intuition, can be thought of as shared access to random coin flips) to try to update them together as often as possible while still respecting the transition probabilities of each respective branch.

Instead of starting from an arbitrary pair of states $x,y \in \Omega$, we apply a simplified version of this proof technique called \emph{path coupling} which was originally proposed by Bubley and Dyer.   Here one defines a path metric $\rho$ on pairs of states in $\Omega$ and shows that an arbitrary pair of states beginning a distance 1 apart with respect to $\rho$ will come closer together (in expectation) after one step of the Markov chain.  

To define a path metric on $\Omega$ first consider a connected graph $(\Omega, E)$ and define $\rho(x,y) = 1$ for each $\{x,y\} \in E$\footnote{Note that in our usage $E$ will be the set of edges corresponding to transitions of $P$, although this is not a requirement in the general method.  In addition, one can assign distinct lengths $\rho(x,y)$ to each edge $\{x,y\} \in E$ but this is not required for our usage.}.  In general a path $\gamma$ in $(\Omega,E)$ from $x$ to $y$ is a sequence $ (\gamma_0,\gamma_1,...,\gamma_r)$ with $\gamma_0 = x$, $\gamma_r = y$, and $\{\gamma_i,\gamma_{i+1}\} \in E$ for each $i$.   From this structure (which is sometimes called a pre-metric) one defines a path metric on the entire set $\Omega$ by
\begin{equation}
\rho(x,y) \equiv \min_{\substack{(\gamma_0,...,\gamma_r) \\ \gamma_0 = x \; , \; \gamma_r = y}} r
\end{equation}
Suppose for each edge $\{x,y\} \in E$ there is a coupling $(X_1,Y_1)$ of $P(x,\cdot)$ and $P(y,\cdot)$ such that
\begin{equation}\label{eq:fastmixcond}
\mathbf{E}_{x,y} \; \rho(X_1,Y_1) \leq \rho(x,y) e^{-\alpha}
\end{equation}
for some $\alpha >0$ then $d(t) \leq e^{-\alpha t} \textrm{diam}(\Omega)$, where $\textrm{diam}(\Omega) = \max_{x,y \in \Omega} \rho(x,y)$, which implies
\begin{equation}\label{eq:mixtime}
t_\textrm{mix}(\epsilon) \leq \alpha^{-1} \log\left(\frac{\textrm{diam}(\Omega) }{\epsilon}\right).
\end{equation}
\section{Proof of rapid mixing}
 Our path coupling applies to the state space graph with vertices $\mathbf{z} \in \Omega = \{-1,1\}^{n\times L}$ and edges $E$ given by pairs of configurations $(\mathbf{z},\mathbf{z}')$ that differ only on a single worldline. If $\{\mathbf{z},\mathbf{z}'\}\in E$ we define $\rho(\mathbf{z},\mathbf{z}')=1$.

Let the initial states $\{\mathbf{z},\mathbf{z}'\}\in E$ differ at a single worldline $i$.  To simulate one step of the Markov chain a worldline $j$ is first chosen to be updated uniformly at random.  In the case that $j$ is not an element of $\mathcal{N}(i)$ then the conditional distributions of worldline $j$, $\pi_j(\,\cdot\,|\mathbf{z})$ and $\pi_j(\,\cdot\,|\mathbf{z}')$, are equal and we can update the worldlines to the same value in the coupling.  Otherwise if $j\in\mathcal{N}(i)$ then $\pi_j(\,\cdot\,|\mathbf{z}) \neq \pi_j(\,\cdot\,|\mathbf{z}')$ due to the influence of worldline $i$ and it is not always possible update the chains to the same value. Therefore the expected distance satisfies
\begin{equation*}\label{eq:expecbound}
    \mathbf{E}_{\mathbf{z},\mathbf{z}'}\rho(X_1,Y_1)\leq 1 -\frac{1}{n}+\frac{1}{n}\sum_{j\in\mathcal{N}(i)}\mathbf{P}_{\mathbf{z}_j,\mathbf{z}_j'}\{X_1^{(j)} \neq Y_1^{(j)} \}
\end{equation*}

By proposition 4.7 in \cite{levin2017markov} \footnote{The theorem states that for any two distributions $\mu,\nu$,
 $\| \mu -  \nu \|_{\mathrm{TV}} = \inf \left \{P(X \neq Y) : (X,Y) \textrm{ is a coupling of } \mu,\nu  \right\}$. } the probability that they are not updated together in the optimal coupling is
 $$ \mathbf{P}_{\mathbf{z}_j,\mathbf{z}_j'}\{X_1^{(j)} \neq Y_1^{(j)}\}  = ||\pi_j(\,\cdot\,|\mathbf{z}) - \pi_j(\,\cdot\,|\mathbf{z}')||_{\mathrm{TV}} .$$
Thus we turn to bounding{\small
\begin{align*}\label{eq:tvd2}
&||\pi_j(\,\cdot\,|\mathbf{z}) - \pi_j(\,\cdot\,|\mathbf{z}')||_{\textrm{TV}}\\
=&\frac{1}{2}\sum_{\bar{z}_j} \left|\pi_j(\bar{z}_j|\mathbf{z}) - \pi_j(\bar{z}_j|\mathbf{z}') \right|\\
=&\frac{1}{2}\sum_{\bar{z}_j} \pi_j(\bar{z}_j|\mathbf{z})\left|1-\frac{\pi_j(\bar{z}_j|\mathbf{z}')}{\pi_j(\bar{z}_j|\mathbf{z})}\right|,
\end{align*}}
First we note that 
$$
\frac{\pi_j(\bar{z}_j|\mathbf{z}')}{\pi_j(\bar{z}_j|\mathbf{z})} = \frac{e^{-\frac{\beta}{L}g_j(\bar{z}_j|\mathbf{z}')}}{e^{-\frac{\beta}{L}g_j(\bar{z}_j|\mathbf{z})}}\frac{\sum_{\bar{z}_j''}e^{-\frac{\beta}{L}g_j(\bar{z}_j''|\mathbf{z})}\phi(\bar{z}_j'')}{\sum_{\bar{z}_j''}e^{-\frac{\beta}{L}g_j(\bar{z}_j''|\mathbf{z}')}\phi(\bar{z}_j'')}.
$$
As the lattice configuration only differs on worldline $i$ we have that
$$
g_j(\bar{z}_j|\mathbf{z'}) - g_j(\bar{z}_j|\mathbf{z}) = \sum_{k=1}^L a_{ij}(z_{ki}'-z_{ki})z_{kj}
$$
from which the bounds
$$
e^{-2\beta a_{ij}}\leq e^{-\frac{\beta}{L}(g_j(\bar{z}_j|\mathbf{z}')-g_j(\bar{z}_j|\mathbf{z}))}\leq e^{2\beta a_{ij}}
$$
follow. As a direct consequence of this we have
$$
e^{-2\beta a_{ij}}\leq \frac{\sum_{\bar{z}_j''}e^{-\frac{\beta}{L}g_j(\bar{z}_j''|\mathbf{z'})}\phi(\bar{z}_j'')}{\sum_{\bar{z}_j''}e^{-\frac{\beta}{L}g_j(\bar{z}_j''|\mathbf{z})}\phi(\bar{z}_j'')} \leq e^{2\beta a_{ij}}.
$$
Putting these together implies
$$
\left|1-\frac{\pi_j(\,\cdot\,|\mathbf{z}') }{\pi_j(\,\cdot\,|\mathbf{z}) }\right| \leq e^{4\beta a_{ij}}-1
$$
since $1- e^{-4\beta a_{ij}}\leq e^{4\beta a_{ij}}-1$.
Therefore
\begin{align*}
&||\pi_j(\,\cdot\,|\mathbf{z}) - \pi_j(\,\cdot\,|\mathbf{z}')||_{\textrm{TV}}\\
\leq& \frac{1}{2}\sum_{\bar{z}_j} \pi_j(\bar{z}_j|\mathbf{z})(e^{4\beta a_{ij}}-1)\\
=&\frac{1}{2}\Big{(}e^{4\beta a_{ij}}-1\Big{)}
\end{align*}
since $\pi_j(\,\cdot\,|\mathbf{z}) $ is normalized.  Summarizing,
\begin{equation}\label{eq:finalbound}
\mathbf{P}_{\mathbf{z}_j,\mathbf{z}_j'}\{X_1^{(j)} \neq Y_1^{(j)} \} \leq \frac{1}{2}\Big{(}e^{4\beta J}-1\Big{)}
\end{equation}
for any worldline $j$, and so
\begin{equation}\label{eq:finalexpecbound}
\begin{split}
    \mathbf{E}_{\mathbf{z},\mathbf{z}'}\rho(X_1,Y_1)\leq 1 -\frac{1}{n}+\frac{\Delta}{2n}\Big{(}e^{4\beta J}-1\Big{)}\\
    \leq \exp\left( - \frac{1}{2n}\left( 2-\Delta\Big{(}e^{4\beta J}-1\Big{)}  \right )\right)\rho(\boldsymbol{z},\boldsymbol{z}').
\end{split}
\end{equation}
as $\rho(\boldsymbol{z},\boldsymbol{z}') =1$.
Setting 
$$ \alpha = \frac{1}{2n}\left[2-\Delta\left(e^{4\beta J}-1\right)\right] $$ in \ref{eq:mixtime} we see that provided $\alpha>0$ we have $t_{mix}(\epsilon) =\mathcal{O}(n\log(n/\epsilon))$, as $\text{diam}(\Omega)=n$. The restriction $\alpha> 0$ is satisfied when
\begin{equation}\label{eq:finalcondbound}
\beta \leq \frac{1}{4J}
\log(\frac{2}{\Delta}+1).
\end{equation}
From $\frac{x}{x+1}\leq \log(1+x)$ one can obtain the weaker but simpler sufficient expression \eqref{eq:thresholdTemp}. From this analysis, a tighter but less transparent upper bound on inverse temperatures that suffice for rapid mixing can be obtained by requiring $\alpha > 0$ for
$$
 \alpha = \frac{1}{2n}\left[2-\max_i\sum_{j \in \mathcal{N}(i)} \left(e^{4\beta a_{ij}}-1\right)\right]
$$
This form is particularly useful when the interaction degree is large, but most of the couplings $a_{ij}$ are small, for example in TIM with qubits embedded in a spatial lattice and interaction strengths $a_{ij}$ that decay as a power law with Euclidean distance.
\section{Concluding Remarks}  These rapid mixing results depend crucially on the use of worldline updates that avoid the critical slowing down of the single-site Glauber dynamics.  In particular, the high temperature regime is said to lead to a ``classical freezing'' (strong ferromagnetic coupling) along each worldline, and the generalized heat-bath updates avoid this by erasing and resampling an entire worldline at once.  This is the reason our result does not follow from known general results on rapid mixing for high temperature classical systems (such systems are known to be rapidly mixing when $\beta J \leq \Delta^{-1}$).  

Since the PIMC method can also be applied to stoquastic Hamiltonians with more general off-diagonal terms than those in \eqref{eq:genTIM}, it is natural to ask whether a similar rapid mixing result holds above some system-size independent temperature for such models (e.g. those containing terms of the form $-X_iX_j$).  The reason our techniques fall short of addressing this case is that $k$-local off-diagonal terms for $k>1$ create strong interactions between worldlines at all temperatures, implying that $||\pi_j(\,\cdot\,|\mathbf{z}) - \pi_j(\,\cdot\,|\mathbf{z}')||_{\textrm{TV}}$ is near 1.  Therefore an extension of these techniques to such cases would likely require considering more general cluster updates than those applied here (such as ``worm algorithm''~\cite{boninsegni2006worm} updates).  

\section{Acknowledgements} 
The research is based upon work partially supported by the Office of the Director of National Intelligence (ODNI), Intelligence Advanced Research Projects Activity (IARPA), via the U.S. Army Research Office contract W911NF-17-C-0050. The views and conclusions contained herein are those of the authors and should not be interpreted as necessarily representing the official policies or endorsements, either expressed or implied, of the ODNI, IARPA, or the U.S. Government. The U.S. Government is authorized to reproduce and distribute reprints for Governmental
purposes notwithstanding any copyright annotation thereon.


\bibliography{bibliography}

\appendix
\section{Worldline Heat Bath Updates}\label{sec:updates}

 To implement each worldline heat-bath we apply an algorithm introduced in~\cite{krzakala2008path} and improved upon in~\cite{farhi2012performance} for sampling the conditional distribution $\pi_j(\bar{z}_j'|\mathbf{z})$ (the algorithm is suitable for any Hamiltonian $H = H_z + H_x$ where $H_z$ is diagonal in the $z$ basis and $H_x = -\sum_{i=1}^n c_i X_i$ where $c_i >0$). For $i = 1,...,L$ define $t = \frac{i\beta}{L}$ and let $z(t) = z_i$.  In order to avoid a dependence on the Trotter number $L$ it suffices to store and track a sequence of events at times $\{ t_1,...,t_m\}$ in which neighboring spins in a worldline flip their value (``jumps''). Notice that the diagonal part of $H$ can be written
$$H_z  = g_j + f_jZ_j$$
where $g_j$ and $f_j$ are operator valued functions involving $\{ Z_k:j\neq k\}$. The influence of the neighboring worldlines on worldline $j$ at imaginary time $t\in[0,\beta]$ can be deduced via $\mathcal{F}_j(z(t)) = \langle z(t) |f_j| z(t) \rangle$. $\mathcal{F}_j(z(t))$ will be a piecewise constant function, switching values whenever a neighbor of worldline $j$ flips its value,
$$
\mathcal{F}_j(z(t)) = 
\begin{cases} 
      h_0& 0 = \tilde{t}_0\leq t\leq\tilde{t}_1\\
      h_1&  \tilde{t}_1\leq t\leq\tilde{t}_2\\
\vdots \\
      h_q&  \tilde{t}_q\leq t\leq \beta=\tilde{t}_{q+1}
   \end{cases}
$$
where the times $\tilde{t}_j$ correspond to times where a neighboring spin has flipped.  

To generate a new path for worldline $j$ the algorithm first computes the value of $\mathcal{F}_j(z(t))$ as a function of imaginary time along the path. Next it generates boundary conditions at the imaginary time points $\tilde{t}_j$ where  $\mathcal{F}_j(z(t))$ changes its value (i.e. a neighboring worldline undergoes a flip). This is done by sampling the distribution
$$
\frac{\langle s_0|A_q|s_q\rangle \langle s_q|A_{q-1}|s_{q-1}\rangle \cdots\langle s_1|A_0|s_0\rangle }{\textrm{Tr}[A_qA_{q-1}\cdots A_0]}
$$
where
$$
A_i = e^{-\lambda_i[h_i Z_j - c_iX_j] }
$$
is a $2 \times 2$ matrix and $\lambda_i = \tilde{t}_{i+1}-\tilde{t}_i$. 

Now that boundary conditions for regions of contstant spin have been chosen, the algorithm generates subpaths on each interval $[\tilde{t}_i,\tilde{t}_{i+1}]$ of length $\lambda_i$. These subpaths are described by a number of flips $w$ and and times $\tau_1,...,\tau_w\in[0,\lambda_i]$ at which flips occur. The number of flips is restricted to being either even or odd depending on the boundary conditions chosen in the previous step. These subpaths are drawn from the probability density on configurations $(\tau_1,...,\tau_w)$
$$
\frac{1}{\langle s_{i+1}|A_i|s_i\rangle}c_j^w\exp\left( -s_ih_i\left[\sum_{j=1}^w (\tau^j -\tau^{j-1}) \right] \right)
$$ 
and finally all subpaths are combined in order to get a full path.

To sample over paths in a region of constant field with fixed boundary conditions, with boundary condition $B_1$ at $t = 0$, the algorithm draws waiting times from an exponential distribution. Starting with $j = 1$, the waiting time $u_i$ until the next flip is drawn from
\begin{align*}
&f(u_i)\\
=& [\sqrt{h^2+c^2}+B_ih]\exp(-u_i[\sqrt{h^2+c^2}+B_ih])
\end{align*}
and set $B_{i+1} = -B_i$, repeating this process repeats until $\sum_j=1^i u_j > \lambda$. At this point the path is output if it satisfies the boundary conditions, and otherwise it is discarded and the process is repeated until success.

\section{Bounding the Time per Update}\label{sec:pfail}
The algorithm described in the previous section has a probabilistic run time.  In order to upper bound this run time and guarantee an FPRAS for the partition function we introduce a failure condition as follows: if at any point during the PIMC method the number of jumps in any worldline exceeds some value $k$ (to be determined below) then terminate and output $0$. We now turn our attention to choosing a suitable $k$. In a given imaginary time region of constant local field $h$ the waiting time drawn for the $i^{th}$ flip is given by
\begin{equation}
f(u_i) = [\sqrt{h^2 + c^2} + B_i h] e^{-u_i[\sqrt{h^2 + c^2} + B_i h]} 
\end{equation} 
where $B_i\in \{-1,1\}$, switching its value after each flip. (i.e. $B_{i+1} = -B_i$). In bounding the probability of a high number of flips we seek the maximum value of $ [\sqrt{h^2 + c^2} + B_i h]$ as when this is maximized the expected interarrival time between each flip is minimized. It is clear that $h_{\text{max}} = \Delta \max_{i,j}|a_{ij}|= \Delta J$ and $c_{\text{max}} = \max_i|\Gamma_i| = \Gamma$, so
$$f_{\text{min}}(u) = [\sqrt{J^2 + \Gamma^2} + J] e^{-u[\sqrt{J^2 + \Gamma^2} + J]} = \lambda e^{-u{\lambda}}$$
where $\lambda = [\sqrt{J^2 + \Gamma^2} + J] $ is the exponential distribution corresponding to the highest rate of spin flips. Physically this corresponds to the neighboring spins being adversarially aligned at each moment in imaginary time. This interarrival process with the fixed maximum rate $\lambda$ corresponds to a Poisson process over the entire imaginary time interval $[0,\beta]$ with the same rate $\lambda$ defined by
$$
\mathbf{P}(x) = \frac{e^{-\lambda\beta} (\lambda\beta)^x}{x!}.
$$
where the random variable $x$ is the number of spin flips in the imaginary time interval.
 Using a Chernoff bound we have:
$$
\mathbf{P}(x>k) \leq e^{-\lambda\beta}e^{-k\left(\log\frac{k}{\lambda\beta}-1\right)}\leq e^{-k\log\frac{k}{\lambda\beta e}}.
$$
Thus the probabililty of observing more than $k$ spinflips decreases exponentially witk $k$ when $k\geq\lambda \beta e$. For simplicity we take $k\geq \lambda\beta e^2$ so that $\mathbf{P}(x>k) \leq e^{-k} $. Since each sample of \eqref{eq:measure} requires $\mathcal{O}(n\log n)$ steps of the Markov chain and assuming whatever quantity we are attempting to estimate requires $\mathcal{O}\left(\textrm{poly}(n)\right)$ samples to compute the probability of failure during entire runtime will obey
$$
\mathbf{P}_\textrm{Fail}\leq c*\textrm{poly}(n)*n\log (n)*e^{-k}
$$
for some constant $c$. We wish to keep $\mathbf{P}_\textrm{Fail}$ below a constant threshold $\mathbf{P}_\textrm{Th} = c*\textrm{poly}(n)*n\log (n)*e^{-k}$. This is equivalent to $k$ obeying
\begin{align*}
 k &=\max\left\{\lambda\beta e^2, -\log\left(\frac{\mathbf{P}_\text{Th}}{c*\textrm{poly}(n)*n\log n} \right)\right\}
\end{align*}
and so $k = \mathcal{O}(\log n)$. Each worldline heat-bath update as described in the previous section therefore runs in time $R=\mathcal{O}(k)$.

\end{document}